\documentclass[%
 reprint,
 amsmath,amssymb,
 aps,
]{revtex4-1}

\usepackage{graphicx}
\usepackage{dcolumn}
\usepackage{bm}
\usepackage{amsmath}
\usepackage{hyperref}
\hypersetup{colorlinks=true,
            breaklinks=true,
            pdfstartview=Fit,
            linkcolor=blue,
            citecolor=blue,
            urlcolor=blue}

\begin{document}

\preprint{APS/123-QED}

\title{Search for axionlike dark matter with a liquid-state nuclear spin comagnetometer}

\author{Teng Wu,$^{1,*}$ John W. Blanchard,$^{1}$ Gary P. Centers,$^{1}$ Nataniel L. Figueroa,$^{1}$ Antoine Garcon,$^{1}$\\  Peter W. Graham,$^{2}$ Derek F. Jackson Kimball,$^{3}$ Surjeet Rajendran,$^{4}$ Yevgeny V. Stadnik,$^{1}$\\ Alexander O. Sushkov,$^{5}$ Arne Wickenbrock,$^{1}$ and Dmitry Budker$^{1,4,6}$}
\affiliation{$^{1}$Helmholtz-Institut Mainz, Johannes Gutenberg University, 55128 Mainz, Germany\\
                     $^{2}$Department of Physics, Stanford Institute for Theoretical Physics, Stanford University, California 94305, USA\\
                     $^{3}$Department of Physics, California State University-East Bay, Hayward, California 94542-3084, USA\\
                     $^{4}$Department of Physics, University of California at Berkeley, California 94720-7300, USA\\
                     $^{5}$Department of Physics, Boston University, Boston, Massachusetts 02215, USA\\
                     $^{6}$Nuclear Science Division, Lawrence Berkeley National Laboratory, Berkeley, CA 94720,USA}

\begin{abstract}
We report the results of a search for axionlike dark matter using nuclear magnetic resonance (NMR) techniques.
This search is part of the multi-faceted Cosmic Axion Spin Precession Experiment (CASPEr) program.
In order to distinguish axionlike dark matter from magnetic fields, we employ a comagnetometry scheme measuring ultralow-field NMR signals involving two different nuclei ($^{13}$C and $^{1}$H) in a liquid-state sample of acetonitrile-2-$^{13}$C ($^{13}$CH$_{3}$CN).
No axionlike dark matter signal was detected above background.
This result constrains the parameter space describing the coupling of the gradient of the axionlike dark matter field to nucleons to be $g_{aNN}<6\times 10^{-5}$ GeV$^{-1}$ (95$\%$ confidence level) for particle masses ranging from $10^{-22}$ eV to $1.3\times10^{-17}$ eV, improving over previous laboratory limits for masses below $10^{-21}$ eV.
The result also constrains the coupling of nuclear spins to the gradient of the square of the axionlike dark matter field, improving over astrophysical limits by orders of magnitude over the entire range of particle masses probed.
\end{abstract}

\maketitle
The identity and properties of dark matter, which makes up over 80\% of the total matter content of the Universe, are still a mystery \cite{Bertone2018}.
While the main evidence for the existence of dark matter comes from its gravitational effects over astronomical distances, the key to discerning its nature lies in identifying its non-gravitational interactions.
The discovery of such interactions would not only illuminate the nature of dark matter but also profoundly impact our understanding of cosmology and astrophysics by unveiling new physical laws and forces \cite{Safronova2018}.

A class of well-motivated dark matter candidates are the canonical (or QCD, quantum chromodynamics) axion and axionlike particles (henceforth generically referred to as axions) \cite{Peccei1977, Peccei1977-2, Rosenberg2015}.
Axions are predicted by theories seeking to explain the strong-CP problem (C and P refer to charge and parity, respectively) \cite{Peccei1977, Peccei1977-2}, the hierarchy problem \cite{Graham2015-2}, and even quantum gravity \cite{Svrcek2006, Arvanitaki2010}.
Axions are spin-0 bosons that can be created in the early Universe via non-thermal mechanisms \cite{Preskill1983, Abbott1983, Dine1983}.
While there are theoretical predictions relating the coupling strength to the particle mass for the QCD axions, the mass of axionlike particles is theoretically unconstrained.
If axionlike particles saturate the observed cold dark matter content, then their de Broglie wavelength must not exceed the dark matter halo size of the smallest dwarf galaxies, giving a lower bond on the axion mass $m_{a}\gtrsim 10^{-22}$ eV.

The large parameter space of axion has motivated many experimental searches based on three possible types of non-gravitational interactions (couplings) between axions and standard model particles: the axion-photon coupling, which can interconvert axions and photons in a magnetic field \cite{Graham2015}; the axion-gluon coupling, which can generate oscillating electric dipole moments (EDMs) in nuclei, atoms, and molecules \cite{Graham2011, Budker2014, Roberts2014, Stadnik2014}; the axion-fermion (wind) coupling, which can induce spin-dependent energy shifts and spin precession in fermions \cite{Stadnik2014, Flambaum2013, Graham2013, Stadnik2017, Graham2018}.
The axion-photon coupling has been searched for in numerous experiments, many of which give constraints for axions with masses heavier than $10^{-6}$ eV \cite{Sikivie1985, Asztalos2001, Ehret2009, Asztalos2010, Brubaker2017, Fu2017, Du2018}.
An optical-cavity experiment was proposed to search for axions with masses of $10^{-17}$ eV and up to $10^{-10}$ eV \cite{Obata2018}.
Recently, the first results from ABRACADABRA-10 cm set upper limits on the axion-photon coupling over the mass range $3.1\times10^{-10}$ eV to $8.3\times10^{-9}$ eV \cite{Ouellet2018}.

It has been proposed that the axion-gluon and axion-fermion couplings could be used to detect axions with masses less than $10^{-6}$ eV by utilizing nuclear magnetic resonance (NMR) techniques \cite{Budker2014, Graham2013, DeMille2017}.
Similar ideas \cite{Stadnik2017} have been applied to analyze experimental data from a search for the neutron EDM, setting the first laboratory constraints for axions with masses ranging from $10^{-22}$ eV to $10^{-17}$ eV \cite{Abel2017}.
The precession frequencies of overlapping ensembles of ultracold neutrons and $^{199}$Hg atoms were simultaneously measured to distinguish the signals from axions and from magnetic field.
Because of the spatial separation between the ensemble-averaged vertical position of the warm $^{199}$Hg and the ultracold neutrons due to Earth's gravitational field, magnetic-field gradients lead to differential effects in the measured spin-precession frequencies of the two species \cite{Baker2006, Sheng2014}.

Here, we report an experimental search for axions using a new comagnetometer configuration.
Different nuclear spins are simultaneously probed within the same molecule in this comagnetometer, which has highly suppressed systematic effects from magnetic-field gradients \cite{Wu2018}.
We use the comagnetometer to perform a month-long search for frequency shifts in nuclear spins induced by axions, and obtain constraints on the axion-nucleon coupling strength for axion masses ranging from $10^{-22}$ eV to $1.3\times10^{-17}$ eV.
Since our work is based on ZULF (zero- to ultralow-field) NMR, and is part of CASPEr (Cosmic Axion Spin Precession Experiment), which is a multi-faceted research program using NMR techniques to search for dark-matter-driven spin precession \cite{Budker2014, Graham2013}, our work is referred to as CASPEr-ZULF-comagnetometer.
Another experiment, CASPEr-ZULF-sideband, is based on the same experimental setup of this work, but with a different search protocol and a different sample.
CASPEr-ZULF-sideband probes axions with masses ranging from $1.8\times10^{-16}$ eV to $7.8\times10^{-14}$ eV \cite{Garcon2018, Garcon2018-2}.
Our experiment, together with the CASPEr-ZULF-sideband experiment, potentially probes the parameter space of axions with masses less than $10^{-14}$ eV.

\begin{figure}[t]
\centering
\includegraphics[width=0.9\columnwidth]{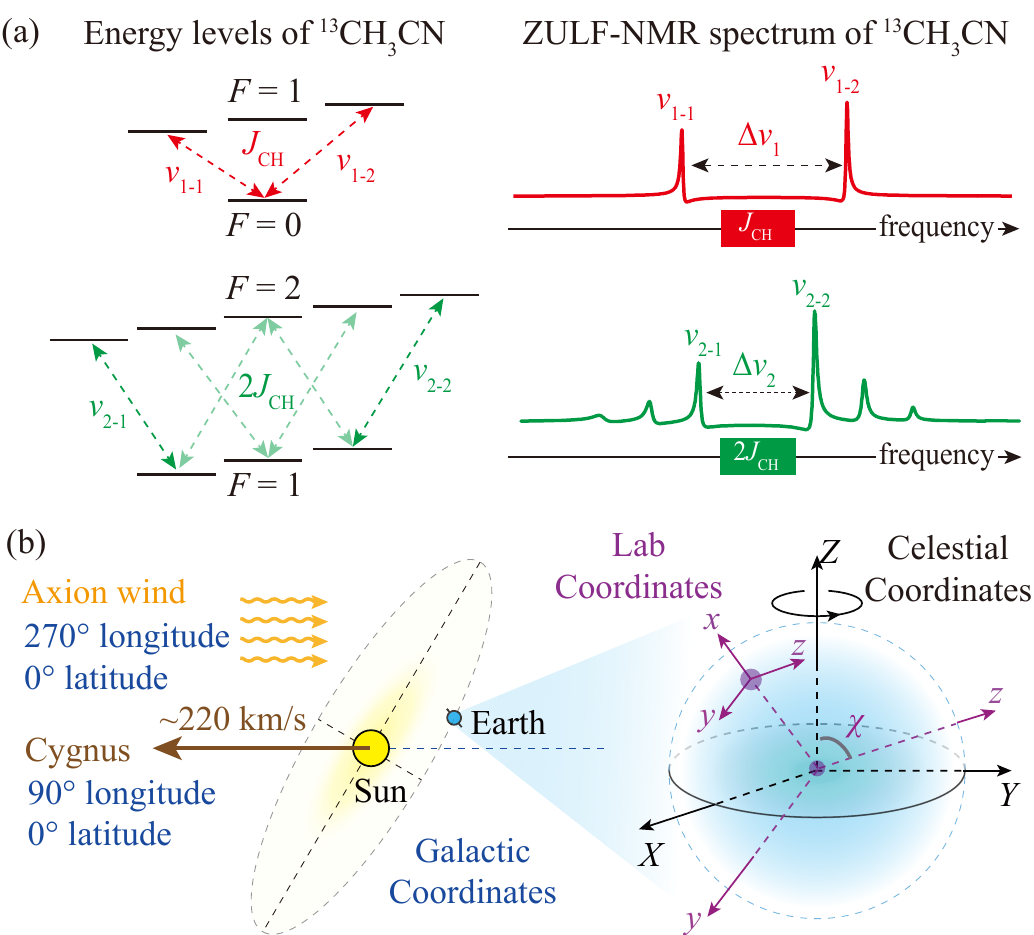}
\caption{(a) The energy level diagram and the ZULF-NMR spectrum of acetonitrile-2-$^{13}$C in a 100 nT bias field. The spectrum has features at frequencies corresponding to $J_{\rm{CH}}$ ($\sim$140 Hz, red) and 2$J_{\rm{CH}}$ ($\sim$280 Hz, green). $\Delta\nu_{1, 2}$ are the frequency splittings used to realize the comagnetometer. (b) The direction of the expected average axion wind velocity. In galactic coordinates, the Sun is moving towards the Cygnus constellation (90$^{\circ}$ longitude and 0$^{\circ}$ latitude). The direction of the expected axion wind velocity is 270$^{\circ}$ longitude and 0$^{\circ}$ latitude. The celestial coordinates are shown in black, and the lab coordinates are shown in purple. For celestial coordinates, the $Z$ direction is the Earth's rotating axis. For lab coordinates, the $z$ axis is the direction of the applied magnetic field. $\chi$ is the angle between $z$ and $Z$ \cite{SI}.}
\end{figure}

The technical details of the apparatus of our nuclear-spin comagnetometer are described in the Supplemental Material \cite{SI}.
The core of our experiment is a ZULF NMR system \cite{Tayler2017, Jiang2018}.
The nuclear spins, $^{13}$C and $^{1}$H in our experiment, are present in a liquid-state sample of acetonitrile-2-$^{13}$C ($^{13}$CH$_{3}$CN, from Sigma-Aldrich, 100 $\mu$L), which is flame-sealed under vacuum in a standard 5-mm glass NMR tube.
The sample is initially polarized in a 1.8-T Halbach magnet, and then dropped into a four-layer magnetic shield (Twinleaf MS-1F). 
The bottom of the tube is thus $\sim$~1 mm above the top of the vapor cell of an atomic magnetometer.
A bias magnetic field is applied using a set of coils within the innermost shield layer in the $z$ direction (laboratory coordinates).
The nuclear spins within the sample evolve under the influence of the indirect nuclear spin-spin coupling ($\it{J}$-coupling), the magnetic field, and the axion field.
After each measurement, the sample is shuttled back into the magnet.
Each individual measurement takes 75 s (with ms uncertainty), including 30 s for data acquisition, with the remaining time for sample prepolarization, shuttling (1 s), and an $\pi$ pulse (50 $\mu$s) \cite{SI}.
The spectrum has features at frequencies corresponding to $J_{\rm{CH}}$ and 2$J_{\rm{CH}}$ \cite{Ledbetter2011, Blanchard2016}.
The measured $J_{\rm{CH}}$ for acetonitrile-2-$^{13}$C is 140.55002(3) Hz \cite{Wu2018}.
In the presence of a small magnetic field, the two peaks split into different patterns, see Fig.~1(a) \cite{Ledbetter2011}.
It is the frequency of the two splittings $\Delta\nu_{1, 2}$ that we use to realize the comagnetometer.
We focus on $\Delta\nu_{2}/\Delta\nu_{1}$, which is insensitive to the magnetic field, but retains sensitivity to frequency shifts induced by axions.

The axions manifest as a classical field oscillating at the axion's Compton frequency $\omega_{a}=m_{a}$, where $m_{a}$ is the axion mass (we adopt natural units, where $\hbar = c = 1$).
Such a field can be written as $a(t)=a_{0}\cos (m_{a}t+\phi)$, where $a_{0}$ is the amplitude of the oscillating field, which can be estimated by assuming that the field energy density $\rho_{a}\approx\frac{1}{2}m_{a}^{2}a_{0}^{2}$ comprises the totality of the local dark matter density $\rho_{\rm{DM}}\approx 0.4$ GeV/cm$^{3}$\cite{Catena2010}.
Here we make a simplifying assumption that $a_{0}=\langle a_{0}\rangle$; however, fluctuations in the amplitude can indeed be important \cite{Gary2019, Derevianko2018, Foster2018, Knirck2018}.
The phase $\phi$ of the local axion field is a random number from 0 to 2$\pi$ of the first measurement.
The coherence time of the axion field observed in a terrestrial experiment is expected to be the duration of $10^6$ oscillations, determined by the virialized velocity distribution of the axions \cite{Kimball2017}.
Including the axion-nucleon coupling, the two frequencies $\Delta\nu_{1, 2}$ of our comagnetometer are \cite{SI}
\begin{align}
\Delta\nu_{1}(\pm) = &(\gamma_{h}+\gamma_{c})B_{z}\\ \nonumber
                       &\pm(g_{app}-g_{ann}/3)\sqrt{2\rho_{\rm{DM}}}\sin(m_{a}t+\phi)v_{z},\\
\Delta\nu_{2}(\pm) = &(\gamma_{h}+3\gamma_{c})B_{z}/2\\ \nonumber
                                 &\pm(g_{app}-g_{ann})\sqrt{2\rho_{\rm{DM}}}\sin(m_{a}t+\phi)v_{z}/2,
\end{align}
where $\gamma_{h, c}$ are the gyromagnetic ratios for $^{1}$H and $^{13}$C, respectively, $B_{z}$ is the magnitude of the bias magnetic field, $g_{app}$ and $g_{ann}$ are the coupling strengths of protons (from $^{1}$H) and neutrons (from $^{13}$C) with axions, respectively \cite{Mayer1950, Klinkenberg1952, Kimball2015}, $\pm$ refers to reversing the magnetic field direction, and $v_{z}$ is the $z$ component of the expected average velocity of the axion wind in laboratory coordinates.

In galactic coordinates, the velocity of the axions with respect to the center of the galaxy is expected to be approximately zero on average, assuming that the standard dark-matter halo is isotropic.
As the Solar system orbits about the galactic center with a velocity of approximately 10$^{-3}c$ and in the direction pointing towards the Cygnus constellation (90$^{\circ}$ longitude and 0$^{\circ}$ latitude in galactic coordinates), the Solar system sees an axion wind in the direction pointing towards 270$^{\circ}$ longitude and 0$^{\circ}$ latitude in galactic coordinates, see Fig.~1(b).
We can neglect the Earth's orbital motion to leading order since the Earth moves around the Sun with a much smaller velocity of about 10$^{-4}c$.
The direction of the expected average velocity of the axion wind in celestial coordinates is thus ($\delta$, $\eta$) $\approx$ (-48$^{\circ}$, 138$^{\circ}$), where $\delta$ is the declination and $\eta$ is the right ascension \cite{Stadnik2017, Abel2017, NASA, Kostelecky1999}.
The $z$ direction in laboratory coordinates should be rewritten in celestial coordinates as well.
Therefore, $v_{z}$ has the form of $[\cos(\chi)\sin(\delta)+\sin(\chi)\cos(\delta)\cos(\Omega_{\textrm{sid}}t-\eta)]|v|$  (detailed conversion between different coordinates is shown in the Supplemental Material \cite{SI}).
Here, $\Omega_{\textrm{sid}}\approx2\pi\times1.16\times10^{-5}$ s$^{-1}$ is the daily sidereal angular frequency, $\chi$ is the angle between $z$ and $Z$, and is calculated as $\cos(\chi)=\cos(\alpha)\cos(\beta)$, $\alpha\approx50^{\circ}$ is the latitude of the Helmholtz Institute of Mainz, and $\beta\approx85^{\circ}$ is the angle between $z$ and North \cite{SI}.
In our experiment, the $z$ direction is parallel to the ground.

\begin{figure}[b]
\centering
\includegraphics[width=0.9\columnwidth]{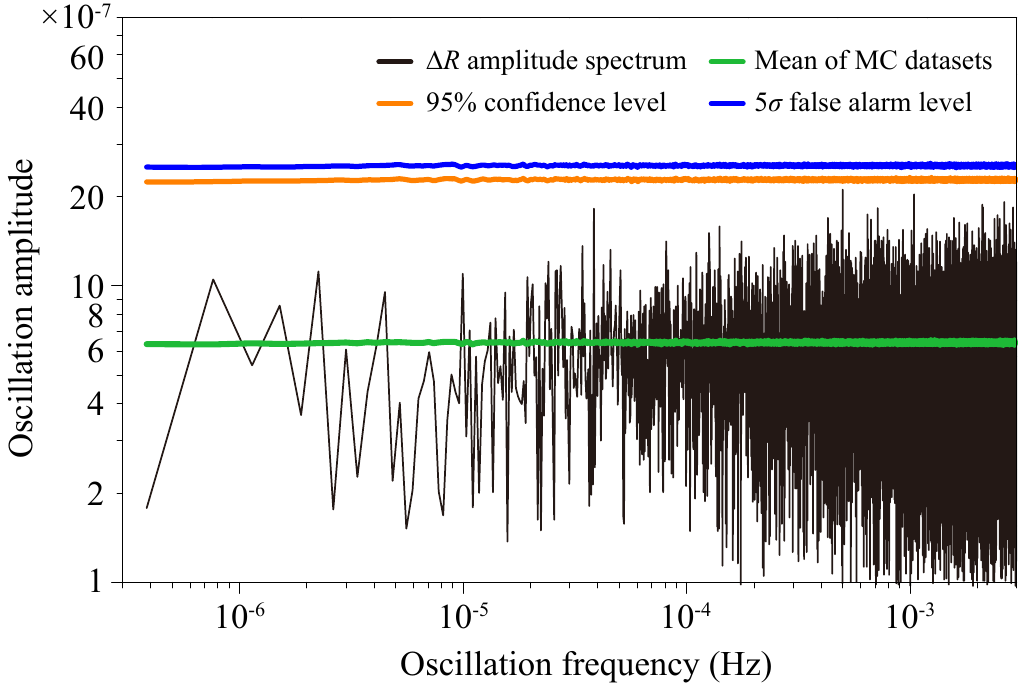}
\caption{The amplitude spectrum of the measured $\Delta\mathcal{R}$ (black line). Under the null hypothesis assuming no axion, the mean of the amplitude spectrum of the collection of Monte-Carlo-generated datasets is shown in the green line. The blue and orange lines depict the false-alarm thresholds corresponding to the amplitude necessary to reach the global $p$ values at 5$\sigma$ level and 95\% confidence level at each frequency, respectively. }
\end{figure}

In order to suppress cycle-to-cycle changes in the magnetic field, we take the ratio $\mathcal{R}$ of $\Delta\nu_{1, 2}$, $\mathcal{R}_{\pm}\equiv\Delta\nu_{2}(\pm)/\Delta\nu_{1}(\pm)$.
The difference in the ratio, $\Delta\mathcal{R}\equiv\mathcal{R}_{-}-\mathcal{R}_{+}$, obtained under field reversal is \cite{SI}
\begin{align}
\Delta\mathcal{R}\approx0.07\sin(m_{a}t+\phi)\cos(\Omega_{\textrm{sid}}t-\eta)g_{aNN}[\textrm{GeV}^{-1}].
\end{align}
We assume that the axion-proton and axion-neutron coupling strengths are the same, i.e., $g_{app}=g_{ann}=g_{aNN}$.
With field reversals, we could suppress systematic effects from the third-order Zeeman effect, drifts in $\it{J}$-coupling strength and chemical shift, which keep the sign in the frequency ratio when the field direction is reversed \cite{Wu2018}.

Based on Eq.~(3), the axion field would manifest itself through time-dependent shifts in $\Delta\mathcal{R}$ at two different angular frequencies: $\omega=|m_{a}\pm\Omega_{\textrm{sid}}|$, which means that a signal generated from the axion field with a much smaller frequency than $\Omega_{\textrm{sid}}$ can be deteced in the higher frequency region around $\Omega_{\textrm{sid}}$.
A third mode with frequency $m_{a}$ is neglected in $\Delta\mathcal{R}$, since its amplitude (proportional to $|\cos(\chi)\sin(\delta)|\approx0.04$) is nearly an order of magnitude smaller than the other two modes (proportional to $0.5|\sin(\chi)\cos(\delta)|\approx0.34$) \cite{SI}.

The measurements were performed for nearly five weeks, from July 14, 2018 (22:42:20) to August 14, 2018 (12:41:35).
Each consecutive $\{+B_{z}, -B_{z}\}$ measurement is used to calculate $\Delta\mathcal{R}$, resulting in a sequence of measurements separated by 150 s.
Occasionally the measurements had to be interrupted for maintenance purposes. 
The precise timings of these downtimes were recorded. 
The black line Fig.~2 is the amplitude spectrum of $\Delta\mathcal{R}$.
In order to determine whether or not there is any signal driven by the axion field, we perform a test with a null hypothesis supposing no axion, and calculate the global $p$ value of the measured signal amplitude (or power) at each frequency.
If the global $p$ value is smaller than a significance level, e.g., 5$\sigma$, we reject the null hypothesis and conclude that there is a significant signal at this particular frequency.

\begin{figure*}
\centering
\includegraphics[width=1.9\columnwidth]{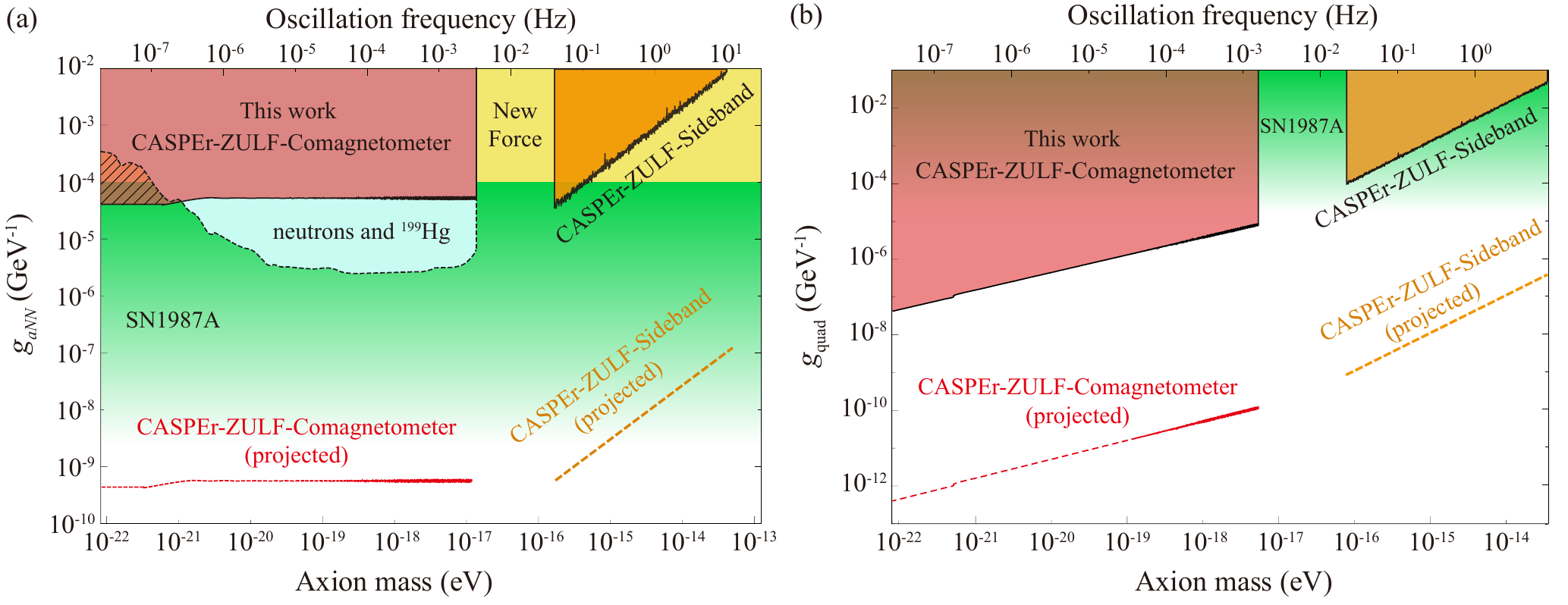}
\caption{Limits on the coupling of nucleons with the gradient of the axion field (a) and with the gradient of the square of the axion field (b). The red region shows the parameter space excluded by this work (CASPEr-ZULF-comagnetometer) at the 95\% confidence level. The region excluded by the PSI neutron EDM experiment is shown in light blue and is surrounded by black dashed line \cite{Abel2017}. The improved region obtained with this work is depicted with black slash lines. The dashed red line shows the projected sensitivity of our system using PHIP (parahydrogen-induced polarization, with a projected 10$^{5}$ enhancement factor) \cite{Theis2012}. The orange region shows the limits given by the CASPEr-ZULF-sideband experiment \cite{Garcon2018-2}, including the projected sensitivity by using PHIP (dashed orange line). Other shaded regions depict constraints from supernova energy-loss bounds (SN1987A, green) \cite{Raffelt1990, Chang2018}, and laboratory searches for new spin-dependent forces (yellow, 95\% confidence level) \cite{Vasilakis2009}.}
\end{figure*}

Under the null hypothesis, we perform Monte-Carlo (MC) simulation to generate a collection of datasets.
For each of the MC-generated datasets, we calculate the power spectrum (the amplitude spectrum shown in Fig.~2 is the square root of the power spectrum).
Based on a number of power spectra from the MC-generated datasets, we obtain the distribution of the power at the $i$th frequency within the frequency range of the power spectrum.
The cumulative distribution function $F_{i}(P)$ of the power $P$ at the $i$th frequency can be fitted with $F_{i}(P)=1-A_{i}\textrm{exp}(-B_{i}P)$, where $A_{i}$ and $B_{i}$ are the fitting parameters.
With these parameters, we can derive the false-alarm thresholds corresponding to different global $p$ values \cite{Abel2017, Scargle1982, Algeri2016}.
See the Supplemental Materials for detailed procedures \cite{SI}.
The alarm thresholds for the 5$\sigma$ level and the 95\% confidence level are shown in the blue and orange lines in Fig.~2, respectively.
Suppose that there is a sufficiently strong axion field whose frequency is within the detectable spectral region, based on Eq.~(3), we should observe two outliers in Fig.~2 in this case, both exceeding the 5$\sigma$ alarm threshold.
The central frequencies of the two outliers should be equal to $|m_{a}\pm\Omega_{\textrm{sid}}|$.
This provides a cross-check method for a true signal produced by axions.
The results in Fig.~2 indicates null detection since the measured signal amplitude at all the frequencies are below these alarm levels.

Following Eq.~(3), we can interpret the 95\% confidence level on $\Delta\mathcal{R}$ (the orange line in Fig.~2) as limits on the coupling of the gradient of the axionlike dark matter field to nucleons $g_{aNN}$.
Due to sidereal modulation, our system can measure axion fields with oscillation frequencies smaller than the frequency resolution, which is around 3$\times 10^{-7}$ Hz and is determined by the total integration time of our experiment.
Such a field would generate a single peak in the amplitude spectrum of $\Delta\mathcal{R}$, with a frequency at $\Omega_{\textrm{sid}}$, since the frequency separation between the two modes $|m_{a}\pm\Omega_{\textrm{sid}}|$ is smaller than the frequency resolution.
The oscillation amplitude at $\Omega_{\textrm{sid}}$ is the sum of the signal amplitude from the two modes and depends on the phase $\phi$ in Eq.~(3).
Considering that $\phi$ is a random value from 0 to 2$\pi$, we take the average value of $|\rm{sin}(\phi)|$.
Thus, the limits obtained from the analysis based on Eq.~(3) have to be multiplied by an additional factor $2/\pi$ if the axion field frequency is smaller than 3$\times 10^{-7}$ Hz (see section 7 of the Supplemental Material for detailed clarification \cite{SI}).

We derive limits on the $g_{aNN}$ and present these limits in Fig.~3(a), shown as the red region.
Compared with the limits given by the PSI neutron EDM experiment (light blue region, surrounded by black dashed line) \cite{Abel2017}, our experiment has improved laboratory constraints for axion masses below 10$^{-21}$ eV, see the red region filled with black slash lines.
The parameter space excluded by the CASPEr-ZULF-sideband experiment is shown as the orange region \cite{Garcon2018-2}.
We note that for the regions excluded by the two CASPEr-ZULF experiments, there still exists a detection gap for axion masses from $1.3\times10^{-17}$ eV to $1.8\times10^{-16}$ eV, corresponding to the oscillating frequency ranging from mHz to tens of mHz.
The lower bound of the gap ($1.3\times10^{-17}$ eV,  $\sim$ 3.3 mHz) is the largest frequency of the oscillation signal that could be probed with our experiment as determined by the time interval of the data (150 s) considering the sampling theorem.
The upper bound of the gap ($1.8\times10^{-16}$ eV, $\sim$ 45 mHz) is determined by the linewidth of the ZULF NMR signal considering the search protocol of CASPEr-ZULF-sideband \cite{Garcon2018-2}.
This gap could be closed with more advanced data processing methods or by reducing the duration of a single measurement cycle.
Other shaded regions depict constraints from supernova energy-loss bounds (green) \cite{Raffelt1990, Chang2018}, and laboratory searches for new spin-dependent forces (yellow, 95\% confidence level) \cite{Vasilakis2009}.

The preceding analysis interprets the CASPEr-ZULF-comagnetometer data in terms of the standard axion wind coupling: the interaction of nuclear spins with the gradient of the axion field.
In some axion models this interaction can be suppressed \cite{Olive2008, Pospelov2013}, in which case the dominant interaction of nuclear spins is with the gradient of the square of the axion field. 
Analogously to Eq.~(3), this quadratic wind coupling to the axion dark matter generates a difference in the ratio $\mathcal{R}$ of $\Delta \nu_{1,2}$ between measurements with opposite applied magnetic field directions:
\begin{align}
\Delta\mathcal{R}\approx1.72\times&10^{-13}\sin(2m_{a}t+\phi)\\ \nonumber
&\times\cos(\Omega_{\textrm{sid}}t-\eta)(g^{2}_{\textrm{quad}}[\textrm{GeV}^{-2}]/m_{a}[\textrm{eV}]),
\end{align}
where $g_{\textrm{quad}}$ is the relevant coupling constant (Eq.~(4) is derived and discussed in the Supplemental Material \cite{SI}).
Constraints on the quadratic axion wind interaction from our measurements are shown in Fig. 3(b), and surpass astrophysical limits by orders of magnitude for all axion masses probed.
Our measurements can also be interpreted to constrain nuclear spin interactions with dark photons \cite{Graham2015, Graham2018} as discussed in the Supplemental Material \cite{SI}.

In conclusion, we have performed a search for axions by monitoring the frequency shifts of nuclear-spin evolution in a liquid-state sample of acetonitrile-2-$^{13}$C.
Our results have shown no significant oscillations driven by axions, and have placed improved laboratory constraints on the coupling strength between nucleons and axions.
A significant enhancement on the detection sensitivity can be obtained by utilizing hyperpolarization techniques to achieve much higher nuclear-spin polarization for the sample. 
Previous work has demonstrated that at least five orders of magnitude enhancement in the signal amplitude can be realized with PHIP (parahydrogen-induced polarization), compared with thermal polarization using a permanent magnet \cite{Theis2011, Theis2012, Suefke2017, Iali2018}.
This will enable our system to search a deeper region of the parameter space (dashed red lines in Fig.~3(a) and 3(b)).
There are several challenges ahead.
Proper catalysts are required to efficiently transfer the singlet spin order from the parahydrogen to the sample \cite{Iali2018}.
Another task is developing procedures to perform continuous non-hydrogenative PHIP \cite{Barbara2018}, which is necessary for a long-term search for axions and other exotic spin-dependent interactions.

We are sincerely grateful to Nicholas Ayres and Christopher Abel for discussions on the Monte-Carlo simulations and to Pavel Fadeev on the axionlike dark matter. This research was supported by the DFG Koselleck Program and the Heising-Simons and Simons Foundations, the European Research Council under the European Union's Horizon 2020 Research and Innovative Programme under Grant agreement No. 695405 (T. W., J. W. B., and D. B.), and by the National Science Foundation under Grant No. PHY-1707875 (D. F. J. K.). Y.V.S. was supported by the Humboldt Research Fellowship. P. W. G. acknowledges support from DOE Grant DE-SC0012012, NSF Grant PHY-1720397, DOE HEP QuantISED award $\#$100495, and the Gordon and Betty Moore Foundation Grant GBMF794.

\end{document}